\documentclass[]{sagej}
\usepackage{lmodern}
 
\usepackage{xspace}

\makeatletter
\newcommand*{\etc}{\@ifnextchar{.}{etc}{etc.\@\xspace}} 

\newcommand{\ie}{i.e.\@\xspace}
\newcommand{\eg}{e.g.\@\xspace}
\newcommand{\cf}{\emph{cf.} \@\xspace}
\makeatother

\usepackage[capitalise, nameinlink, sort]{cleveref}

\usepackage{xcolor}
\usepackage[most]{tcolorbox}

\usepackage{todonotes}
\usepackage{url}

\newcounter{example}[section]
\renewcommand{\theexample}{\arabic{example}}

\newcommand{\example}[1]{%
  \vspace{0.2cm}%
  \refstepcounter{example}%
  \begin{tcolorbox}[colback=gray!10, colframe=gray!40, boxrule=0.5pt, left=2pt, right=2pt, top=2pt, bottom=2pt]
    \textbf{Example \theexample.}~#1
  \end{tcolorbox}%
}

\begin{document}


\author{Muffy Calder\affilnum{1},
Marion Oswald\affilnum{2}, 
Elizabeth McClory-Tiarks\affilnum{3},
Michele Sevegnani\affilnum{1} and
Evdoxia Taka\affilnum{1}}

\affiliation{\affilnum{1} School of Computing Science, University of Glasgow  \\
\affilnum{2} School of Law, University of Northumbria \\
\affilnum{3} School of Law, University of Newcastle}

\runninghead{Calder et al}

\corrauth{Muffy Calder, School of Computing Science, University of Glasgow, Glasgow G12 8RZ, UK}
\email{muffy.calder@glasgow.ac.uk}

 \newpage
 
 \title{Responsible AI in criminal justice:  LLMs in  policing and risks to case progression}

 \begin{abstract}

There is growing interest in the use of Large Language Models (LLMs) in policing, but there are potential risks.   
We have developed a practical approach to identifying risks, grounded in the policing and legal system of England and Wales. 
We identify 15 policing tasks that could be implemented using LLMs and 17 risks from their use, then illustrate with over 40 examples of impact on case progression.
As good practice is agreed, many risks could
be reduced. But this requires effort: we need to address these risks in a timely manner and  define
  system wide impacts and benefits.



 \end{abstract}

\maketitle

\section{Introduction}\label{sec:intro}

The rapid growth of Artificial Intelligence (AI) over the last decade \cite{Thormundsson2025} has included  significant advances in the field of foundation models --  models trained on massive and diverse data sets to perform a wide range of tasks.  Specifically, Large Language Models (LLMs) 
  are  foundation models   trained on vast text corpora,   becoming ``one of the fastest – if not the fastest – adopted technologies in history'' \cite{ELONUNINewsBureau2025}.
Policing is no exception as a domain for  this technology adoption.  There is growing  interest  in  the use of LLM-based tools and increasing availability of  in-house and commercial    tools, and in some countries, e.g. some states in USA, there is emerging interest in regulation \cite{Conversation,SB524,SB180}. 


%

This motivates us to  identify and  understand the risks of introducing LLM-based tools.  Specifically, 
the use of AI in this context brings  risks
in terms of   {\em case progression} -- the management of a legal case as it moves through the criminal justice system.  Obvious and well-known examples include hallucinations, omissions, and biases, but we propose there are more risks and their  impacts on case progression can be very specific to this context.    

While assessing software generated outputs in the UK criminal justice system is not a new problem (\cf~post-office horizon system~\cite{Bates}), 
  there is a need for clear guidance in the context of AI. 
This is  indicated by the recent (UK) Ministry of Justice call for evidence~\cite{MoJ} and the Criminal Justice Joint Inspectorate report on Joint Case Building by the Police and  Crown Prosecution Service (CPS)~\cite{Jointcase}, which   calls  for improved handling and management of cases between the police and the CPS and mitigations for the risks of using AI. 
While there is    high level guidance for policing (e.g.~\cite{covenant} in the UK), and on broader  concepts such  as   human rights, data protection, privacy, and right to a fair trial ~\cite{Selcuk}; 
we perceive a current gap concerning  specific, practical risks to case progression, which  all tool users need to take seriously.

 We aim to address that gap and the overarching research question we address is: 
{\bf what are the possible risks  and adverse impacts of deploying LLM-based tools on case progression in England and Wales?}

We focus on risks that might have an adverse impact on case progression, i.e.  adversely impacting decisions so that a case progresses when it should not; or does not progress when it should. For example, an LLM might add content to a witness statement, making an otherwise weak case appear stronger. This could lead to a decision to charge, when otherwise the result would be no further action - so the case might progress, when it should not. Similarly, an LLM might add content to an otherwise strong case and if this addition is identified by the defence and challenged, it may cast doubt on the reliability of the witness and could undermine the case - so  a case may then end up \textit{not }progressing, when it should.



  While grounded in  policing in England and Wales, our findings may be relevant in other jurisdictions. 
  
We address the research question in two steps.     

First, we identify a variety of policing tasks where the tools could be used, or experimentation is already taking place in some police forces, across several stages of  the criminal justice system. The time frame for potential use of these tools is the near future, \ie next five years.  Example tasks include policy retrieval, intelligence summarisation, statement  transcription and translation, form filling, and statement generation.  Productivity and general administrative tasks are not considered.  We   consider features and capabilities that might be involved, rather than  specific tools. 

Second,  we identify  potential, practical risks of  using these tools in policing and impacts on case progression. The examples are grounded in the legal system in England and Wales.

\subsection{Research contribution}
This is a first of its kind systematic study of the risks and impacts of LLM-based tools on case progression.   Addressing these risks  will not only improve  case management in the context of AI, but reviewing them could also serve as a valuable guide for defence lawyers.        The risks can   also guide future regulation and  technical developments, \eg towards tools that are less likely to impact negatively on case progression.

Specific research contributions are: 
 
\begin{itemize}
    \item Identification of 15 example policing tasks that could be implemented using LLM-based tools, across four stages of the criminal justice system. These are plausible  scenarios  that could be adopted within the next five years,   many are already in the  proof of concept or  trial stage~\cite{WP1}.

   \item Categorisation of risks  of using LLM-based tools for these tasks,   relating to the tool outputs, inputs, evaluation, and systems engineering. 
 
\item Identification of 17 risks, distributed across the four categories. 

\item     42 example impacts,  arising from the identified risks and tasks,  illustrating that the risks are not abstract concepts but  can have real-world  impacts on  case progression.   
\end{itemize}

\subsection{Organisation of paper}

Background on AI and related work is given in   \cref{sec:related}, and a  framework that identifies 8 stages of the criminal justice system   is outlined in   \cref{sec:CJ}.  This is relied on to consider the potential for impact on case progression beyond the policing stages.
  The example policing tasks are defined in \cref{sec:tools} and the risks, with example impacts, are   outlined  in 
  \cref{sec:risks}. 
Discussion of the significance of the risks  follows in  \cref{sec:discussion}, 
with  conclusions in \cref{sec:conclusions}. The Appendix   contains a list of the risks.




\section{Background and related work}\label{sec:related}

\subsection{From language models (LMs) to large language models (LLMs)}

Language models are  models trained on text data (in an unsupervised way, i.e. without providing any labels for the text training data) for natural language processing tasks, especially language generation. They have evolved significantly over the last two decades leading to four generations of models \cite{zhao2025}. 
The first generation is   pure task-specific statistical models (i.e. n-gram models \cite{Lee2025}) for (next or missing) word prediction. 
The second generation is models that exploit neural networks (i.e. a special type of machine learning model, which trains layers of artificial neurons to perform a task similarly to how the brain does - thus, called a deep learning model) to create word representations for task-agnostic feature learning (e.g. Word2Vec \cite{Johnson2023} is a commonly used such model). 
The third generation is models that exploit transformers \cite{Vaswani2017} 
to create context-aware word representations, and are pre-trained on vast datasets (e.g. generative pre-trained transformers (GPTs) like GPT 1 and 2). These models allow task-transferable problem solving through fine-tuning (i.e.  adapting the pre-trained model for a specific task through further training on a smaller dataset that directly reflects this task \cite{Bergmann2024}).  
The fourth generation is 
 ``general-purpose task solvers'' large language models (LLMs) with prompt-based user-model interaction to solve a wide range of real-world tasks (e.g. GPT 3/4, Claude).

This transition from simple language models (LMs) to scaling large language models (LLMs), and the increasing task solving capacity of language models~\cite{zhao2025}, are attributed to the emergence of new techniques (e.g. transformers) and the increase in the size (i.e. number of parameters) of models. The prompt interfaces enable the general user to interact with the LLMs usually to achieve a desired task.

A prompt is user-provided text-based input  for the language model \cite{Ggaliwango2024}. It is usually phrased as a query or command to achieve the desired tasks. It can contain longer statements to describe the context (e.g. a character for the LLM to mimic), provide instructions, or refer to the conversation history. The prompt might also contain multimedia (images, videos, or audio) or any other type of input data and query something based on them. Examples of expected outputs can be also included in the prompt, and depending on the number of examples  provided,  we differentiate between zero-, one-, and few-shot prompting~\cite{Ramesh2025}. Prompt engineering refers to the process of writing the prompt, e.g. how to phrase the query, specify the style, choose the words, grammar, contextual details, examples etc. to include, and     strategies for effective prompting that lead to high-quality and relevant outputs \cite{Geroimenko2025}. 

We expect most of the tools used in policing to be based upon commercially available, general and  pre-trained LLMs. These   could be  fine-tuned
 for the policing domain, though   such fine-tuning is rare (one  exception is academic research on predictive policing using   data in the United States~\cite{smartpolice}). Currently, LLMs, with their increasing size, can rival the fine-tuned, pre-trained language models in performance for various complex tasks simply through (zero- or few-shot) prompting~\cite{guo2023}.

 LLMs provide a 
broad linguistic foundation for tasks such as 
text generation and question answering, but it is important to acknowledge  that their reasoning is fundamentally
distinct from human and symbolic logical inference. Thus, they 
 can generate misleading or
factually incorrect content, commonly referred to as “hallucinations”~\cite{LLM_hallucination}, and they may struggle to maintain logical consistency throughout an extended discourse~\cite{kumar2025}. 

A recent technique that blends LLMs with information retrieval is sometimes used to mitigate the hallucination issue. Namely, Retrieval Augmented Generation (RAG) connects LLMs to external knowledge sources (repositories)  through  retrieval mechanisms, allowing an LLM to access and utilise additional information in the repository.   Responses to prompts may   include enumeration of the sources (\ie citations to documents) and there is a  prioritisation of content in the repository over trained content, thus allowing LLMs to be {\em augmented} by new content without re-training. 
 While RAG improves accuracy and reduces risk of hallucination, the latter is still possible in RAG, for example, due to (document) retrieval failure  or generation bottleneck~\cite{LLM_hallucination}.

 
 

\subsection{LLMs beyond text generation}

Language models traditionally are text-to-text models, namely models that take text as an input, process it and generate some new text as an output. LLMs could be used for any text-related task like text summarization, language translation, conversational agents (e.g. chatbots), literature review etc. LLMs can also be used as an extra modality within multi-modal systems (i.e. a system that involves different types of data or methods) with the purpose to enhance the performance of the overall system. There are increasing scenarios where LLMs can be used for pre- or post-processing.  For example, LLMs could be used in automatic speech recognition (ASR) tools: the transcripts generated by the speech-to-text ML model are inserted to a fine-tuned LLM to fix errors, e.g. in spelling, grammar etc. or speaker diarisation (i.e. allocations of text to speakers) \cite{peng2024,Wang2024,Efstathiadis2025}.

The newest generation of AI models consist mainly of \emph{multimodal LLMs}, i.e. models that are capable of processing and generating multiple types of data, e.g. text, audio, image, video by exploiting the power of pre-trained core LLMs. The multimodal LLMs convert the inputs of different data type including the user-provided text instructions into numerical representations called embeddings and then align these into a common space. This allows a core LLM model to analyse and understand the relationships between diverse inputs and answer questions about multimodal inputs or generate multimodal responses. Multimodal LLMs can perform tasks beyond a language focus, e.g. image or audio captioning, charts explanation, information searches in videos, optical character recognition, spoken responses generation, image generation from text description etc. \cite{Wu2025,deitke2025molmopixmo,dai2024nvlm}. 

\subsection{Related work}



While there is increasing awareness of generic risks of using LLMs (\eg as mentioned in \cref{sec:intro})  
 there appears to be little work focussed  on LLM-based policing tools, especially in  the context of case progression.
 A notable exception  is the  evaluation~\cite{Walley}  of the  Hertfordshire Constabulary proof of concept {\em ADA} tool that employed LLMs to  generate a  MG11 police  witness statement form, which        documents witness accounts of events in a structured and formal manner.
 The evaluation is based on 
  user and stakeholder interviews, focussing on  analysis of  the quality and 
linguistic style of the AI output and  productivity and process performance.  The  analysis is based on a comparison between the quality of reports produced by  humans and the tool  (though not on reports generated from the same set of data). 
The evalution contains useful 
definitions of quality and 
insights into hallucinations, which we will discuss  further in 
\cref{sec:risks}.


\section{Criminal justice context}\label{sec:CJ}

The wider context of the criminal justice process is useful to consider, particularly as LLM-based tools used for policing tasks could have an impact at later stages of the process, e.g. affecting the admissibility or weight of evidence adduced at trial. There are a number of different bodies and stages which make up the Criminal Justice System (CJS) in England and Wales,
and no single model of the system and components; here we adopt the following framework, breaking the process down into 8 stages: (1) Community policing and offender management; (2) Intelligence; (3) Investigation; (4) Charging decision or alternative disposal; (5) Trial or guilty plea; (6) Sentencing; (7) Prison and parole; (8) Probation.  

The first three stages involve policing and are where the LLM-based tools could be used directly; they may be used indirectly  in stages 4 and 5, through influencing decision-making at these stages, and/or constituting evidential material.   \textbf{Community policing and offender management} involves the police co-operating with community groups and multi-agency approaches targeted at tackling persistent offenders and/or offenders assessed as being at high risk of reoffending. \textbf{Intelligence} is the use of information given to or collected by police to initiate or support ongoing investigations. The \textbf{Investigation }stage is where police collect and consider evidence to investigate the alleged offence, e.g. collecting witness statements or risk and threat assessments, interviewing the suspect, considering any digital (e.g. data from computers or phones) or forensic (e.g. DNA) information as potential evidence.
 The \textbf{Charging decision }can be made by police for less serious offences, or police can offer an \textbf{alternative disposal}, such as a caution or penalty notice. The Crown Prosecution Service (CPS) makes the charging decision for more serious offences. This involves two stages: the \textit{e}\textit{vidential stage}, i.e. consideration of whether there is enough admissible and reliable evidence for a realistic prospect of conviction; and the \textit{public interest stage}, i.e. consideration of whether a prosecution is required in the public interest, taking into account factors such as the seriousness of the crime and the suspect’s age and maturity.

 Where a decision to charge a person is made, the case will go to court for an initial hearing. All cases involving an adult defendant start in the magistrates’ court\footnote{We will focus on the process for adult defendants. There are differences in how youth court trials are run and different alternative disposals and sentencing options available for defendants aged under 18.}. More serious cases will be sent to the Crown Court, but over 90\%  of all criminal cases will be concluded in the magistrates' court.

 If the defendant enters a \textbf{guilty plea}, the case will then move on to the \textbf{sentencing }stage. If the defendant pleads not guilty, there will be a \textbf{trial} held in the magistrates’ court or Crown Court, depending on the seriousness of the offence. In the \textbf{magistrates’ court}, there is no jury and decisions on questions of both law and fact are made by the district judge or lay magistrates hearing the case. In the Crown Court, decisions about law are made by the Crown Court judge and decisions about fact, including whether the defendant is guilty or not, are made by the jury. This means that where issues of law arise during a trial process, the jury can be sent out and a decision is made by the judge in their absence, \eg about the admissibility of evidence. In contrast, a district judge or lay magistrates will make these decisions. Where they make a decision that evidence is inadmissible during the course of a trial, they have to disregard it when making a decision as to guilt in the magistrates' court, but could still be influenced by having seen or heard that inadmissible evidence. 
 

If the defendant is found guilty, the case will move to the \textbf{sentencing }stage. This may not be on the same day, depending on whether there is more information needing to be gathered about the defendant, e.g. a pre-sentence report (which will include a risk assessment) prepared by probation. In the magistrates’ court, the district judge or lay magistrates will decide on sentence; in the Crown Court, the judge will decide on sentence – the jury plays no part in sentencing.

There are a number of different sentences available, including \textbf{prison}, in which case a decision will need to be made about the category of prison that the defendant should initially be sent to. This involves assessing risk and may make use of a previous risk assessment carried out by probation for any pre-sentence report.
When a prisoner becomes eligible for \textbf{parole}, a decision will be made about whether they are able to be released on licence (this also involves an assessment of risk) – in which case they will be managed in the community by probation. 

Crimes committed in prison can be dealt with by the Prison Adjudication System, where an adjudication process – which is similar to a trial, but with fewer procedural safeguards – may be held to determine guilt and decide punishment. Punishments include additional days imprisonment, cell confinement or loss of privileges. Alternatively, a more serious crime may be referred to the police. A protocol is in place for determining how a crime committed in prison should be dealt with \cite{HandlingCrimesInPrisonAgreement}.

Where a defendant is sentenced to a non-custodial sentence, they may require management and supervision by \textbf{probation} in the community, for example where they are subject to a community order with particular requirements, e.g. unpaid work or electronic monitoring.

We note there could be a number of different contexts and reasons why the progression of a case could be adversely impacted at different stages of the CJS, i.e. might not progress when it should; or might progress when it should not. For example, a case might not progress at the intelligence
stage due to a piece of information being missed or misinterpreted, and at the trial stage, a key witness's credibility might be undermined by inaccuracies in their witness statement, undermining the case and leading to an acquittal.
Misinterpreted information at the intelligence stage or inaccuracies contained in a witness statement could also lead to a case progressing when it should not, where these errors make a case appear stronger than it is. This could lead to decisions to progress an investigation or charge a suspect, when no further action should have been taken.



\section{Policing tasks that could involve LLM-based tools}\label{sec:tools}

 We identify 15  policing tasks that could be implemented using LLM-based tools, across the first four CJS stages.  
 The  list is not  exhaustive,  but  is indicative of  features and capabilities
  that  could be adopted within the next five years.   It is informed by  literature reviews and research on mapping the current use of AI tools in UK policing~\cite{WP1}.  Examples  of some tools are   already in the  proof of concept or  trial stage. 
 We have tried to avoid repetition   therefore while some tools/tasks could appear in more than one stage, we have only mentioned them once.

  


\subsection {Community policing
and offender management}

\begin{itemize}
 \item   A {\em policy aid} for officers -- answering policy queries  from policing policy and regulation  repositories/ knowledge bases (\eg PACE\footnote{PACE (Police and Criminal Evidence Act 1984)     regulates police powers and protects public rights in the UK.} codes of practice, etc.)
 with {\em retrieval-augmented generation (RAG)}. Such a RAG   could be mobile, \ie accessed in real-time on a hand-held device.

 \end{itemize}

 \subsection{Intelligence}
 There are several tasks that involve  summarisation, search, or redaction.    
 \begin{itemize}
 \item   {\em Document} summarisation  of one or more (text) reports. 
  \item   {\em Visual representation}  summarisation, for example using an LLM to summarise trends in a graphic.    
 \item   {\em Video} summarisation, for example  using LLMs to translate video frames   from body worn cameras into sequences  of captions and then assess relative importance of each frame.  
 \item{\em Video} search interface  for frames matching a description of a person or object, for example, search for ``person on a bicycle''. 
 \item{\em Redaction} of text from documents, objects or text from images,  or frames from video.
 \end{itemize}

\subsection{Investigation}\label{Inv}

 There are several tasks  concerning witness/victim interviews and statements.  These     are more fine grained than those listed for the previous two stages, and consequently they are more numerous.  This reflects  both the wide variety of uses of text in investigations and the breadth of a number of current proof-of-concepts trials, though   note that  some tasks, \eg transcription and translation, may also occur in the previous two stages.  

\begin{itemize}
 \item {\em Transcription} of  an audio witness/victim interview  to  text.    
 
 \item {\em Text translation} of a (text) witness/victim interview  transcript in one natural language to English.

\item    {\em Question generation}  for a witness or victim,  in the context of a specific offence. 

\item   {\em Generation of a witness/victim statement} in chronological order (\eg for a MG11), from an (text) interview transcript.  



 \item  {\em Key word generation}  from a witness interview statement or summary.
 
   \item   {\em Corroborations generation} from   a set of (text) witness statements, i.e. corroborating text from different statements. 

 \item  {\em Generation of the points to prove} that have not been addressed in  a witness statement, for a specific offence.   Points to prove are   key to charging decisions and are specific to the offence.     
 
 \item  {\em Generation of  answers to risk assessment questions}  in a specific  template (\eg DARA\footnote{Domestic Abuse Risk Assessment tool https://www.dashriskchecklist.com/}), or from  a (text) witness interview transcript or statement.

\end{itemize}


\subsection{Charging decision}

The main task concerns the generation of reports. 
\begin{itemize}
 \item      {\em Generation of a police report} \eg a  MG5, which  
  summarises the details of a case and is prepared in advance of the  charging decision.
This would form the basis of the prosecution in the event of a guilty plea.
\end{itemize}

\section{Risks}\label{sec:risks}

There are several known categorisations of risks.  Three  relevant examples include  the  technical risk factors, individual harm risks factors, systemic risk factors of the \emph{AI in our
Justice System, 2025} report~\cite{JUSTICE1} created by JUSTICE, a UK charity organization;
the  SSAFE-D Principles (Sustainability, Safety, accountability, Fairness, Data Stewardship) and  Context-Based Risk Assessment (COBRA) from the  Alan Turing Institute  \emph{Guide to ethical and responsible design, development, and deployment of Artificial Intelligence, 2023}~\cite{ATI}; and the  five cross-sectoral principles set out in the \emph{UK Government’s 2023 AI White Paper}~\cite{Whitepaper}: safety, security and robustness; appropriate transparency and explainability; fairness; accountability and governance; contestability and redress. But none of these are helpful in addressing our question: what are the possible risks and impacts of deploying LLM-based tools {\em on case progression}?  
In other words, this is a systems issue: what is the impact of introducing a new {\em tool}, at a given stage,  on this and subsequent stages in a {\em process}?     

\begin{figure} 
\begin{center}
    \includegraphics[scale=.35]{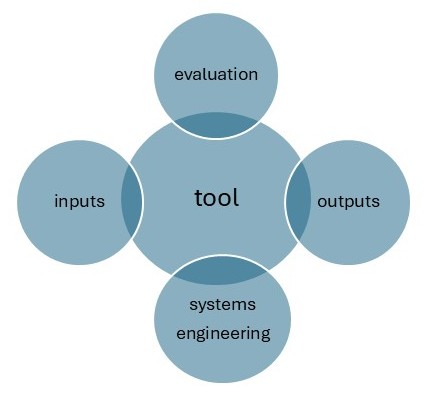}
    \caption{Four categories of risk of deploying a LLM-based tool: inputs, outputs, evaluation, and systems engineering.}
    \label{diagram}
    \end{center}
\end{figure}

We have therefore proposed    
four categories of risks relating to tool   {\em outputs},  {\em inputs},   {\em evaluation}, and   {\em systems engineering} (see Fig.~\ref{diagram}).  
These   categories could apply to the introduction of any tool (whether or not AI) because we have focussed  on what we can {\em observe}:   the outputs and inputs to the tool, how to evaluate whether the tool is fit-for-purpose, and how the tool has been engineered.

We have identified     
seventeen risks.
Many of the risks are well-known consequences of properties of LLMs -- properties that are intrinsic (\ie inherent to the model itself) or extrinsic (\ie dependent on external   context   or inputs, e.g. training data),     
but others have been identified because of a specific relevance in the   criminal justice system.  In each case, we  focus on aspects that could be relevant to case progression, rather than more high level, generic  concepts such as human rights, data protection or privacy implications.  Not surprisingly, we find there are substantially more risks associated with the outputs category, compared with the other categories.

For each risk, we  give an outline   and then one or more   examples relating to the tasks mentioned above and a description of   potential impacts. For some examples, we give a suggested mitigation.  The risks are repeated, as a simple list, in the Appendix.




\subsection{Outputs}\label{results}

\subsubsection{Completeness}\label{completeness}
 This is one of the most substantial risks and  refers to  the potential for  (erroneous) {\em omissions} or {\em additions} in the outputs that have been generated. 
 We note that in the context of statements,  identifying when information has  been  omitted or overlooked, or added by a ``hallucination'',    relies, in part,  on the memory of the interviewer and participant(s).   
On the other hand, we also note that omissions, in particular, are possible even when there is no use of AI, so the use of an automated tool may motivate greater attention to the possibility.   

 
Omissions or additions produced by an LLM-based tool could weaken an otherwise strong case, or strengthen an otherwise weak case. Where such information feeds into charging decisions, it could lead to otherwise weak cases progressing where a suspect should have had no further action taken against them; and where a case is weakened by invented or misleading information, it could lead to a suspect being undercharged (charged with a lesser offence), or not being charged at all.
Once a charging decision has been made and the defence have been served evidence relating to a case, omissions or additions contained in that evidence might influence the advice defence barristers or solicitors provide to their clients about the strength of the case against them – potentially affecting decisions about going to trial or pleading guilty.

Case progression could be impacted at the trial stage where omissions or additions from an LLM-based tool are contained in a witness's statement. This is clearly the case where erroneous additions find their way into the final witness statement, but even if errors are identified and rectified prior to the statement being finalised, the version with the LLM’s ``invented'' evidence should be included as a draft on the schedule of unused material. It could therefore be relied on by the defence to suggest that different information was initially provided by the witness and therefore cast doubt on the reliability of the witness’s evidence.

 \example{
 ``Hallucinations'' may be additions, and one such addition was uncovered   in a generated witness statement from the {\em ADA} proof of concept tool (reported in~\cite{Walley}, page 38). The generated statement included the (hallucination) text “she is a woman of average height and brown hair”, which  was a clear addition because {\em no} description of the woman was given during the interview. }

\example {If a witness or victim uses  a profanity, this   could be  omitted from a transcription, a translation, or summary,  as a result of    LLM safeguards or guardrails~\cite{LLM_safe}.}

\example{ Poor context and design of a  prompt (or prompt conversation),  could impact text or video summarisations  or corroborations generation, leading to the omission of information at the intelligence stage.}

\vspace{.2 cm}

These  examples  could impact significantly on the integrity of evidence  and/or influence directions taken in the intelligence and investigation stages.
However,  an omission might  have little impact if it occurred during use of a tool that  provides a new capability. This is illustrated by the following example. 

 \example{
 There is a risk that as a result of a video search, e.g. for ``person on a bicycle", some relevant frames are not returned.  But, there may be such a huge volume of video data that   without automation, they could not be searched within a reasonable timeframe, e.g. it might take several weeks to carry out the search manually. So, the return of {\em any} frames could be considered to be favourable to   the alternative, which is  no frames at all.}


\subsubsection{Accuracy}  
 This is another substantial risk and it refers to generated content being {\em correct}. There may be several causes of this risk and it may be difficult to identify the cause, depending on the task involved. The following examples illustrate the range of   potential causes.   
 
 \example{ There may be   a   ``factuality'' hallucination~\cite{LLM_hallucination},  these are instances where the LLM generates information that is factually incorrect or not supported by (real-world) evidence.  Examples include references to events that did not occur in a generated witness statement,  or to a legal precedence that did not occur. This may impact    the integrity of evidence and/or legal process.

A recent, real-life and high profile example of the impact of inaccuracies on police decision-making  is the decision to ban Maccabi Tel Aviv fans from a football match in 2025.  Some of the intelligence that informed this decision was subsequently    revealed as inaccurate \cite{inspection}, and the case attracted national media attention   \cite{BBC}.  A letter from the West Midland Chief Constable   \cite{letter}, states ``the erroneous result concerning the West Ham v Maccabi Tel Aviv match arose as result of a use of Microsoft Co Pilot… [we had] understood that the West Ham match had only been identified through the use of Google.''
}\label{accuracy}    

\example{Errors may  result from      
 ambiguity or poor context and design of a  prompt (or prompt conversation), which could impact text or video summarisations  or corroborations generation, leading again to issues with the integrity of evidence.}  

\example{ A translation tool may erroneously replace or rephrase offensive language in  a witness or victim statement, which may impact on the integrity of evidence.}       

\example{Using an LLM in a post-processing step to fix errors in wording or speaker diarization in transcripts generated by automatic speech recognition tools can lead to a higher number of (new) errors \cite{peng2024} if the LLM is not fine-tuned \cite{Efstathiadis2025,Wang2024}. }

\vspace{.2 cm}


As with omissions or additions, inaccurate content generated by an LLM-based tool could affect the integrity of charging decisions and also adversely impact the advice given to defendants about the case against them.
Where inaccuracies find their way into witness statements, then - as with omissions and additions - this may result in the witness's credibility being undermined during cross-examination. Note that a barrister has a duty to promote their client's interests, so far as that is consistent with the law and their overriding duty to the court. In the adversarial system in England and Wales, it is part of the role of a defence barrister to promote the defendant's case and undermine the prosecution case. This will involve identifying inaccuracies and inconsistencies in the prosecution evidence and cross-examining the relevant prosecution witness accordingly.

\subsubsection{Timeliness of training and repository  data}
 
LLMs learn from data sets that represent knowledge available at the time of their creation.  A risk is that the content or emphasis in generated outputs may be outdated.

\example{ A (RAG) policy aid may include out of date information, which   could result in breach of PACE Codes of Practice. While this does not \textit{automatically} mean exclusion of evidence, it could potentially lead to a stay of proceedings for abuse of process (i.e. criminal proceedings are stopped); civil/criminal proceedings brought against police; or disciplinary proceedings against officers.}

\example{ There is a risk that video summarisation may order the relative importance of frames according to outdated information (\eg that does not include recent, related terrorist events). This could influence intelligence directions.}  
 
\example{ A visual representation summarisation may  be outdated or not reflect current best practice, e.g. the use of an inappropriate  infographic        could influence intelligence directions.}

\example{  Question generation, in the context of specific offence, may not take into account recent legal outcomes or guidance, impacting the   integrity of evidence.}

\subsubsection{Linguistic style}
These risks range from basics such as poor grammar and spelling,  to more sophisticated concepts such as poor fluidity, fluency, coherence, and readability,  and the imposition of  possibly inappropriate or unauthentic  styles.     

\example{  LLM safeguards  may remove     offensive language   (see  also \ref{completeness}) and thus   reduce  fluency and readability, and thereby impact integrity of evidence. The precise wording used may be essential to establishing an element of the offence, e.g. an assault, or to establishing other important information bearing on the offence, e.g. things said by the suspect on arrest, or what was said by a police officer when arresting a suspect - which may have particular relevance in certain cases, such as where a defendant is charged with resisting a police constable in the execution of their duty.}

\example{ Where there are multiple police officers at the scene of an incident, their raw  witness statements can often be formulaic and  similar. 
After processing by an LLM, there is a risk they could  become   (near) identical, providing defence advocates with cross-examination material, namely,  comparing
two officers’ statements about the same incident and suggesting collusion rather than  independent versions of events. }

\vspace{.2 cm}

\example{ Any generated content may impose   styles that obscure, minimise, or  remove human emotion and sentiments~\cite{zhang2024}, impacting the integrity of evidence and direction of intelligence.  
For example, using a LLM as an interview post-processing step may 
 fail to retain and reflect a victim's emotional tone of speech during their interview.}

\example{ As raised above in relation to completeness and accuracy, there is a risk that a witness’s evidence in chief (the witness's own evidence about the case) is markedly different to the (generated) witness statement. If so, defence barristers could cross examine the witness on this point. This is done to undermine the credibility of the witness and may have adverse implications on the trial if sufficient doubt over the witness's evidence can be raised.}

 \subsubsection{Cultural Language Misinterpretations}\label{cultural}

LLMs have been found to struggle in translation with idiom- or dialect-related misinterpretations attributed to cultural differences \cite{Singh2024,Yao2024}, this presents a risk to any task that involves witness input, e.g. speech or text.

\example{  Summarisation and transcription tools may not have been trained on  local dialect and colloquialisms, and so  there is a risk that some  phrases may be  incorrectly, or at best inappropriately, represented in the generated output.  For example, references to individuals, their roles, illegal goods or activities could be  omitted, obscured, or incorrectly transcribed,  which  may affect the quality of intelligence and also the investigation.}   

\example{ 
The same tools may not be able to represent completely or accurately individuals with  particular communication difficulties, with  mental health or learning difficulties,   neurodiverse witnesses/defendants, child witnesses/defendants, or  speakers of English as a second language  ~\cite{HunterBharda,LeVigne}.}

\vspace{.2 cm}

These risks 
may affect the credibility of the witness, which  could be undermined if they give evidence at odds with what is in their witness statement at court.  
Cases  
involving witnesses with communication difficulties are particularly concerning, given the potential for the intersection of language needs with other vulnerabilities, e.g. individuals being victims of human trafficking.


\subsubsection{Social and cultural biases}


Biases embedded in the training data 
may  give rise to social and/or cultural biases in the generated output. 
For example, 
the way  texts are selected to formulate the training corpora of LLMs can  be a source of selection/sampling bias   \cite{Navigli2023},   leading to social bias in the generated text (e.g.  gender, sexual and racial biases, or bias related to minorities and disadvantaged groups).  

\example{There is a risk of gender bias in the pronouns appearing in generated witness statements:  certain professions or individuals with particular roles may be referred to more frequently with a masculine (or feminine) pronoun~\cite{Kotek2023}.  This might not be noticed by the humans at the statement review stage.}

 \example{There is a risk of difference of emphasis in summaries, depending on the gender of the subject.  This has been shown (with respect to gender) in health care record summaries~\cite{Rickman2025}. }



\vspace{.2 cm}

 Victim blaming language is an ongoing concern within policing \cite{VBL}.
 
 \example{If the corpora used for training the language models contains victim blaming language, or officer statements contain such language, there is a risk that outputs could replicate this language.      } 

\vspace{.2 cm}

There are already known problems with the quality of risk assessments and pre-sentence reports prepared by Probation, both generally \cite{probation}
 and particularly for ethnic minority offenders 
\cite{probation_race}.  The use of LLMs therefore has the potential to exacerbate these existing problems, possibly   leading to inaccurate risk assessments and inaccurate decision-making based on those assessments of risk.

\example{ There is the risk to bias in  offenders' risk assessments when LLMs are used to fill in risk assessment forms, e.g. DARA, by either biasing the scores or changing the emphasis depending on the protected characteristics of offender.}
\vspace{.2 cm}
 There are well-publicised cases where inaccurate risk assessments have been identified as causes in an offender going on to commit serious further offences whilst under supervision by probation in the community (\cite{Bendal} is an example). 

\example{A risk assessment that is too high   could feed into a sentencing decision to imprison someone who should not be imprisoned (with adverse resource implications and an impact on the individual, their family and any dependents); a risk assessment that is too low  may result in the release into the community of someone who should be imprisoned.  
}



\subsubsection{Intermediate, unused material}\label{intermediate}
Tools may produce outputs   that are considered  as drafts, either produced at different stages of the tool or before and after  human edits. They may also  generate  new types of data such  as meta data. But there is currently no consensus on how these outputs should be treated, \eg required to be recorded, stored, and listed on the schedule,  and consequently there is a risk to the integrity of legal processes and  more work. 

\example{ There is a risk of  additional time and resource implications for disclosure to the CPS.  This was encountered by the Hertfordshire Constabulary  during the ADA proof of concept trial:  “Another consequence is the need to disclose all versions
of the transcript and statement produced by officers when using ADA. This impacts on workload for the Disclosure
officer and the CPS.” \cite{Walley}. This can have an impact on `case velocity', i.e. the time it takes for a case to progress.} 

\vspace{.2 cm}

As has been widely-reported, there are currently significant delays in the criminal justice system, with a record number of outstanding Crown Court cases \cite{Justice_Delayed}. The risk of tools adversely impacting case velocity adds to the problem of ``delay and inefficiencies in the progression of cases [which] can have a significant impact on victims, witnesses and defendants.'' \cite{commitment}. 


\example{  There may be many iterations in the process to generate an output, there is a risk that the length of time or the number of steps taken, to agree a final draft  suggests police
influence on witness’s account.}

\example{There is a risk that in the event of evidence deemed inadmissible in magistrates’ courts, district judges or lay
magistrates will still have read the material and thus be influenced by it.}

\subsection{Inputs}

\subsubsection{Prompt and prompt conversation design and effectiveness}\label{prompt}

 A significant aspect of all LLM-based tools is the role of prompt(s), whether they are constrained (\ie pre-coded in a template for the user) or not.  Specifically, prompt design   affects outcomes, for example,   accuracy and compliance~\cite{Promptdesign2025}, 
and   biases, nuances,  and stylistic aspects in prompts can skew the results~\cite{Geroimenko2025,ChatGPTEuropol}. 
 Furthermore, this extends to  the whole iterative process of prompting, and the resulting ``conversation''  -- the list of user prompts and LLM responses.  When prompts are not pre-coded, there will be questions concerning    who is formulating the prompts,  is that pre-determined,  and  are such decisions  recorded, as well as how have  prompts and conversations  been recorded and how are they  accessed.

\example{  If all the information concerning prompt formulation and prompt conversations   is not available and consistent with expected practice, there is a risk that a court may question whose {\em version} of events a generated statement {\em tells}, for example, that of  a police officer or  a young/vulnerable witness. }

\example{ Prompt conversations can be considered a form of intermediate material and  as such there is risk to the workload and to  integrity of the legal  process, similar to that described in \cref{intermediate}.}

 \subsubsection{Prompt   attacks}\label{promptattacks}
 
 The prompt engineering interface for an LLM  
is a ``largely underestimated attack surface'' \cite{Janowski2025}. 
There are at least four emerging adversarial prompt engineering techniques that can be considered as \emph{prompt attacks}.  First, a \emph{prompt injection} attack   can   override system prompts (\ie pre-coded developer instructions) in agentic setups,  with the aim of changing the model output or issuing a tool command. This is especially dangerous when the model is connected to external tools/APIs to e.g. send emails, query databases, or execute code. Second, \emph{jailbreaking} attacks are prompts that   bypass safeguards with the aim of generating harmful or biased content, for example,  
  instructing the model with the ``Do Anything Now (DAN)'' prompt,  to bypass ChatGPT's safeguards \cite{Shen2024,ChatGPTEuropol}. 
  Third, \emph{prompt extraction} attacks can reveal system prompts with the aim of extracting details about the business logic, safeguards, or architecture,  and fourth, \emph{information leakage} attacks   can make the model diverge from its dialogue-style generation to memorisation, with the aim of revealing sensitive data or proprietary information injected in the training data or stored user conversations/prompts and memorized \cite{Janowski2025,ChatGPTEuropol,Nasr2023}. 

Attacks such as these pose   risks to the integrity of legal processes, similar to cyber-security concerns.  They can also pose specific risks such as revealing sensitive information about witnesses and evidence, as well as risks to  tool integrity and reliability (see also \cref{reliability}).

\example{There is a risk a prompt injection attack could enable a user to bypass the constraints of pre-coded prompts and thus influence an output in ways not intended by tool, nor expected by users of the outputs. }

\example{ There is a risk a prompt extraction attack   could enable access to the information categories that an LLM-based automatic redaction system aims to   redact. Then,  through a prompt injection or a jailbreaking attack,  a user could intervene   and change or remove  the information categories,  thus altering  or avoiding redaction.}

 \subsubsection{Leakage to third parties}
Information can be leaked, either intentionally or unintentionally.
Key risks include    exposure of sensitive information, personally identifiable information, proprietary information, system prompts, details of business logic, safety measures, or architecture to unauthorised third parties (e.g. users, attackers) \cite{ChatGPTEuropol,Janowski2025}. For example, sensitive information could be exposed when sensitive data in training datasets is memorized and reproduced in outputs, or injected in stored user prompts \cite{BAHUGUNA2025,Nasr2023}, or during real-time interactions due to misconfigured APIs, or weak access controls \cite{BAHUGUNA2025}. 


\example{There is a risk that  exposure of sensitive information could lead to issues being identified with evidence associated with LLM tool use, which could affect the fairness of proceedings. This could result in evidence being excluded under  section 78 Police and Criminal Evidence Act (PACE) 1984 \cite{PACE} -- this provision allows a judge to exclude prosecution evidence where ``it appears to the court that, having regard to all the circumstances, including the circumstances in which the evidence was obtained, the admission of the evidence would have such an adverse effect on the fairness of the proceedings that the court ought not to admit it."}

\example{ More generally, leakage may be a data protection issue and so there is a risk that additional time and effort has to be  spent managing  breaches, thus delaying case progression.}


\subsection{Evaluation}
LLMs    pose  new challenges,  especially when compared with machine learning classification tools where  there 
 is ground truth and  established statistical measurements for evaluation (such as precision and recall).

\subsubsection{Lack of ground truth}\label{ground}
 
The evaluation of   results that involve language,  such as summarisation, may be highly subjective and there is  no ground truth with which to compare, \eg what is a ``good'' summary.  As mentioned earlier, this is quite different from machine learning for classification, where there is ground truth. The lack of ground truth may  pose a risk to any subsequent  interpretation or decision making  that depends on these   outputs. 

\example{ After  a meeting or interview and  a summary is generated, there is a risk that participants might disagree on whether or not something was discussed and/or the conclusion as stated in the summary is reached. In the absence of a form of recording of the actual meeting, there is no agreed ground truth to refer to -- only individual   memories.}

\subsubsection{Testing process and benchmarks}
 
A significant risk, and major research challenge, 
 is  a lack of consensus on  evaluation criteria, testing processes,  and benchmarks for LLM-based tools in policing.   There remain open questions on   how they should be defined, and by whom,  and subsequently, testing and evaluation results  may be inconsistent, incomplete, incomparable, and/or subjective. A recent systematic review of 445 LLM benchmarks showed that  many were not valid measurements~\cite{bean2025measuring}. 

While  many evaluation techniques used for   traditional Natural Language Processing (NLP) are applicable to LLMs, these cannot deal with many of the LLM-specific issues raised in \cref{results} such as hallucinations or evaluating high-level semantic qualities \cite{Schaik2024}. Furthermore, the evaluation mode, \emph{human} or \emph{automated}, can affect the evaluation criteria. For example,
 when evaluating the linguistic style of  generated text, human evaluations are generally more focussed on readability, fluidity, and similarity to a human generated “gold standard”, whereas automated metrics are more likely focussed on coherence, consistency, fluency, and relevance \cite{Schaik2024}. 
 
\example{There is a risk that LLM-based translators are not accredited, and so statements or evidence translated by an LLM-based tool does not have the same standing in court as those translated by an individual accredited by the    UK Police Approved Interpreters and Translators Scheme (PAIT) (introduced by the National Police Chief's Council (NPCC)).}

\example{There is a risk that   evaluation benchmarks are not updated \cite{peng2024}, so the  data they contain may become outdated.  Furthermore  the benchmark data may leak and thus contribute to training data (resulting in overfitting). This would compromise the integrity of an accreditation process, and subsequently the evaluated tools.}  


\example{ LLMs as evaluators is a new direction for automated evaluation, \ie using one LLM to evaluate (the outputs of) another. Many of the  risks and impacts mentioned above will apply, especially since LLMs suffer from biases \cite{peng2024}.}

\subsection{Sytems Engineering}

\subsubsection{Tool reliability}\label{reliability}


This risk concerns  whether the tool is trustworthy, which usually involves showing it performs {\em consistently}.   Demonstration of reliability depends on testing regimes, access to prompts and conversations,  and  it  may also be related to version and configuration control.  Note that reliability cannot  be interpreted simply as  repeatability, since LLMs are stochastic: the same input to an LLM may result in a different output at different times (e.g. today's output compared with tomorrow's). Therefore, a
statistical framework is required. Further, it may be relevant to distinguish between consistency of tool {\em performance},   given same inputs, and consistency of tool {\em use}, \ie consistency of inputs.    

\example{Unconstrained prompt and prompt conversations may be a risk to   demonstrating  both consistent  tool performance, because as mentioned in \cref{prompt}, prompt design affects outcomes, and tool use, because each user may interpret the need for a prompt differently.     Both  may affect  credibility of the tool, which in turn affects the credibility of its outputs. }   

 \vspace{.2 cm}

To minimise this type of risk, constraints are often added to   the prompts \eg they are pre-coded.


\subsubsection{Tool version/configuration}\label{version}
Tools may be  updated \eg with new functionality or improved  performance,  and before using any version, it may be necessary to configure it  by setting various parameters.  

\example{There is a risk that tool version and configuration settings may not have been recorded, or  the version used for a particular task may no longer be available, thus it may not be possible to repeat,  compare, or confirm the exact date of any generated results. }

\subsubsection{Tool supplier/vendor}
 Tools may be in-house or  developed and  supplied by a third party.

\example{There is a risk that a tool, or information concerning it, may no longer be available, because the supplier/vendor is no longer trading, or simply does not respond to requests.} 

The two examples above could affect case progression, where a lack of available information limits the extent to which the reliability of a tool can be demonstrated. This could, for example, undermine evidence associated with the tool at trial, where the operation of the tool is challenged.

\subsubsection{Tool chaining}
If LLM-based tools are {\em chained}, \ie the outputs of one are the inputs of another, then the   cumulative effect of all the risks may be difficult to qualify or quantify. Moreover, it may not be transparent when and how chaining has occurred, especially if the chain involves more than one workflow or organisation.  

\example{ A simple example of chaining that might not be transparent is  where police officers have used   a translation tool, such as Google translate,  
in the process of   employing an LLM to generate the witness statement~\cite{HunterBharda}.}
 
A more nauanced example is the following.

\example{
 The case summary could be relied on at  sentencing hearings where a guilty plea has been entered and
no trial has taken place (note that the latest MOJ \cite{Courtstatistics} figures 73 per cent of magistrates' court and 61 per cent of Crown Court cases resolve in a guilty plea). There is a risk that inaccuracies and “invented” evidence from the witness statements, which make their way into
a case summary, could therefore form the basis of the prosecution’s submissions to the sentencing judge or magistrates,
with resulting sentencing decisions based on incorrect information and sentences being either too lenient or too harsh
for the actual offending behaviour. In some cases, this information will also feed into risk assessments and potentially affect decisions about the offender's suitability for certain types of sentence, prison category, or later decisions about parole.}


\vspace{.2 cm}


\subsubsection{Expert witnesses}
It may be unclear {\em who} can be an expert witness to explain a tool and what they would say, 
 including  the ways that  risks are identified and mitigated. There is no consensus about what details about inner workings of a tool would be required.  
 
\example{ There is a risk that tool experts with the requisite skills are not available or they are not able to understand what is needed or communicate effectively with non-experts.}


\section{Discussion}\label{sec:discussion}


We have considered the risks, and their impacts,  of general (AI) technology (LLMs)  
in a very specific domain-- policing and criminal justice.

Some risks and their impacts may be considered {\em fundamental}, such as completeness, accuracy, and   sensitivities to  prompt design.  They are fundamental in   two senses:  they are intrinsic to this type of AI and they  could  adversely impact case progression.  We will never be able to eliminate these risks, but over time, we expect technological improvements so that the likelihood of their occurrence is reduced, and we may be able to   anticipate and mitigate them. But the use of language will always remain  subjective. Particularly in the context of LLM as pre- or post- processors, developers and users should question why the pertinent information cannot be presented in a raw format, instead of within a text, which will, to an extent, be interpretive. 
Further, as Example \ref{accuracy}  shows, the use of an  LLM instead of a search engine  introduces a  significant  risk  with potentially  far reaching, and public,  impacts.

Other risks, such as those associated with the systems engineering, may be regarded as {\em transient}.  In time, we expect   codes of practice will be developed  that will address most, if not all of these risks.   This also applies to processes that currently   incur higher  workloads  because of lack of consensus,  such as recording, storing, and putting information on the schedule (\cref{intermediate}), and leakage to third parties. 
We also expect the   risks to  guide future technical developments, e.g. towards tools that are less likely to impact negatively
on case progression. 
 
In all of the risks, the   role of humans is crucial, for example,  to review and mediate. But there are also  fundamental  risks related to the nature of human interactions that will remain, such as relying on individual memories   (\cref{ground}) and  the example involving interpretation of the number of steps taken  (\cref{intermediate}). 

We hypothesise that risks associated with translation  may be  ``easier''   to mitigate, through the use of benchmarks. This is because the task is well-known (within and beyond policing) and has been performed for a long time, well before NLP and LLMs.  Thus,  measures of performance evaluation may be more established,  though this is still a topic of research. 
We also propose that RAGs,  for areas such as policy that depend crucially on current and up-to-date knowledge,  present  lower levels of risk, when document repositories are kept  up to date.

 
We note there will be 
 heightened risks for some parts of the CJS, where there are reduced due process protections, including limited
opportunity for legal representation and  closed conditions. For example, Prison Adjudications and the Single Justice
Procedure in the magistrates,  where a single magistrate decides a case on the papers in a closed court
(see \textit{Northern Trains
Ltd v Ballington and Others} [2024] 8 WLUK 114  \cite{Ballington} for some of the problems which have arisen with this procedure).


Finally,  we remark that in this paper we have focussed on risks\footnote{We have in effect produced a crib sheet for defence lawyers!}.    We have paid  little attention  to    potential   benefits, especially   uses that could  address the current,  huge, court backlogs. However, even when beneficial, there may be additional work introduced into the system, not least because the use of LLMs will require more scrutiny, in the short term.  
More broadly,  to articulate benefits, requires i)  articulating what problem  the use of LLM technology is supposed to address, and ii)   demonstrating the problem has been addressed effectively and efficiently.  And to answer that,   requires systems thinking, namely,  what is the impact of introducing a new {\em tool}, in this case an LLM-based tool,  into a {\em process}, in this case,  the criminal justice process. This paper specifically centres on and articulates this systems thinking.


\section{Conclusions}\label{sec:conclusions}

There is growing interest in the use of LLM-based tools in policing,  with increasing availability of in-house commercial tools.  In the near future (i.e. within five years),  there are many policing tasks that  could employ LLM-based tools. But, there are risks, with potential impacts on case progression.  

 The aim of this paper has been to identify and illustrate those risks and impacts.  We have adopted a practical approach, considering how case progression might be affected not just at policing stages, but at later stages of the criminal justice system too, including at trial and sentencing.  We defined  a set of example policing tasks that could be implemented using LLM-based tools,  and then identified  17 risks arising from the use of these tools and  over 40 examples of how they could have real-world impacts on  case progression.  
The   risks are grouped as relating to outputs, inputs, evaluation, or systems engineering, and we have considered how those risks might prevent a case from progressing to the next stage when it should progress, as well as when a case could be impacted so as to progress when it should not. This latter consideration is of particular pertinence considering the huge backlog of criminal cases and overcrowded prisons, to ensure resources are directed appropriately and to avoid miscarriages of justice. While grounded in policing in England and Wales, many of the risks will  be relevant in other jurisdictions.

LLM technology is changing quickly, and also the role of LLMs, especially  as interfaces to other tools.   With time, we expect that many of the risks, and examples, described here  will be reduced -- as the technology evolves and {\em good practice}, \ie  standards, rules, and enforcement and oversight of these, is agreed. Though of course new risks may be introduced by new capabilities.  But this will not happen without effort.  If we are going to use these tools in the near future,  we need to address the {\em present} risks, seriously, and in a timely manner. 

Finally, we reiterate there must be clarity about {\em why} an LLM-based tool is being introduced, and what are the system wide impacts and benefits.


 
\newpage
 \section*{Appendix}\label{sec:appendix}

Risks, as  identified in \cref{sec:risks}.

\vspace{.4cm}

\begin{tabular}{l l }
{\bf Category} &  {\bf Risk} \\
\hline
Outputs &1 Completeness  \\
 &  2 Accuracy \\
 & 3 Timeliness of training and repository data\\
 &4 Linguistic style\\
 &5 Cultural Language Misinterpretations\\
 &6 Social and cultural biases\\
 &7 Intermediate, unused material\\
 \hline Inputs
&8 Prompt design and conversations\\
&9 Prompt attacks\\
&10 Leakage to third parties\\
\hline Evaluation
&11 Lack of ground truth\\
&12 Testing process and benchmarks\\
\hline Engineering
&13 Tool reliability\\
&14 Tool version/configuration\\
&15 Tool supplier/vendor\\
&16 Tool chaining\\
&17 Expert witnesses
 
\end{tabular}





\bibliographystyle{plain}
 

\end{document}